\documentclass[twocolumn,prb]{revtex4}
\usepackage{amsfonts}
\usepackage[T1]{fontenc}
\usepackage{amsmath,amsbsy,amssymb,graphicx}
\usepackage{times}

\let\mathbf=\boldsymbol

\begin{document}

\title{Electric circuits for non-Hermitian Chern insulators}
\author{Motohiko Ezawa}
\affiliation{Department of Applied Physics, University of Tokyo, Hongo 7-3-1, 113-8656,
Japan}

\begin{abstract}
We analyze the non-Hermitian Haldane model where the spin-orbit interaction
is made non-Hermitian. The Dirac mass becomes complex. We propose to realize
it by an $LC$ circuit with operational amplifiers. A topological phase
transition is found to occur at a critical point where the real part of the
bulk spectrum is closed. The Chern number changes its value when the real
part of the mass becomes zero. In the topological phase of a nanoribbon, two
non-Hermitian chiral edges emerge connecting well separated conduction and
valence bands. The emergence of the chiral edge states is signaled by a
strong enhancement in impedance. Remarkably it is possible to observe either
the left-going or right-going chiral edge by measuring the one-point impedance.
Furthermore, it is also possible to distinguish them by the phase of
the two-point impedance. Namely, the phase of the impedance acquires a dynamical
degree of freedom in the non-Hermitian system.
\end{abstract}

\maketitle

\textit{Introduction:} Non-Hermitian topological systems open a new field of
topological physics\cite%
{Bender,Bender2,Malzard,Konotop,Rako,Gana,Zhu,Yao,Jin,Liang,Nori,Lieu,UedaPRX,Coba,Jiang,JPhys}. 
They are experimentally realized in photonic systems\cite%
{Mark,Scho,Pan,Weimann}, microwave resonators\cite{Poli}, wave guides\cite{Zeu}, 
quantum walks\cite{Rud,Xiao} and cavity systems\cite{Hoda}. The
winding number\cite{Kohmoto,Fu,Yin} and the Chern number\cite{Kohmoto,Fu}
are generalized to non-Hermitian systems. There are some properties not
shared by the Hermitian topological systems\cite%
{Bender,Bender2,Malzard,Konotop,Rako,Gana,Zhu,Yao,Jin,Liang,Nori,Lieu,UedaPRX,Coba,Jiang,JPhys}. 
For instance, the Hall conductance\cite{Chen,Phil,Hirs} and the chiral
edge conductance\cite{WangWang} are not quantized in the non-Hermitian Chern
insulators although the Chern number is quantized\cite{Fu}. The typical
model of the Chern insulator is the Haldane model, but there is so far no
extension to the non-Hermitian model in literatures. Furthermore, although
there are several studies on non-Hermitian Chern insulators\cite%
{Kohmoto,Fu,Yao2,Kunst,KawabataChern}, there are so far no reports on how to
realize them physically.

In this paper, we propose to realize the non-Hermitian Haldane model by an
electric circuit. An electric circuit is described by a circuit Laplacian.
Provided it is identified with a tight-binding Hamiltonian\cite{TECNature},
any results obtained based on a tight-binding Hamiltonian may find
corresponding phenomena in an electric circuit. Indeed, the SSH model\cite{ComPhys}, 
graphene\cite{ComPhys,Hel}, Weyl semimetal\cite{ComPhys,Lu},
nodal-line semimetal\cite{Research,Luo}, higher-order topological phases\cite%
{TECNature,Garcia,EzawaTEC}, Chern insulators\cite{Hofmann}, non-Hermitian
topological phases\cite{EzawaLCR,EzawaSkin} and Majorana edge states\cite{MajoTEC} 
have been simulated by electric circuits. The edge states are
observed by measuring the impedance\cite%
{TECNature,ComPhys,Garcia,Hel,Hofmann,EzawaTEC,EzawaLCR,EzawaSkin}.

\begin{figure}[t]
\centerline{\includegraphics[width=0.48\textwidth]{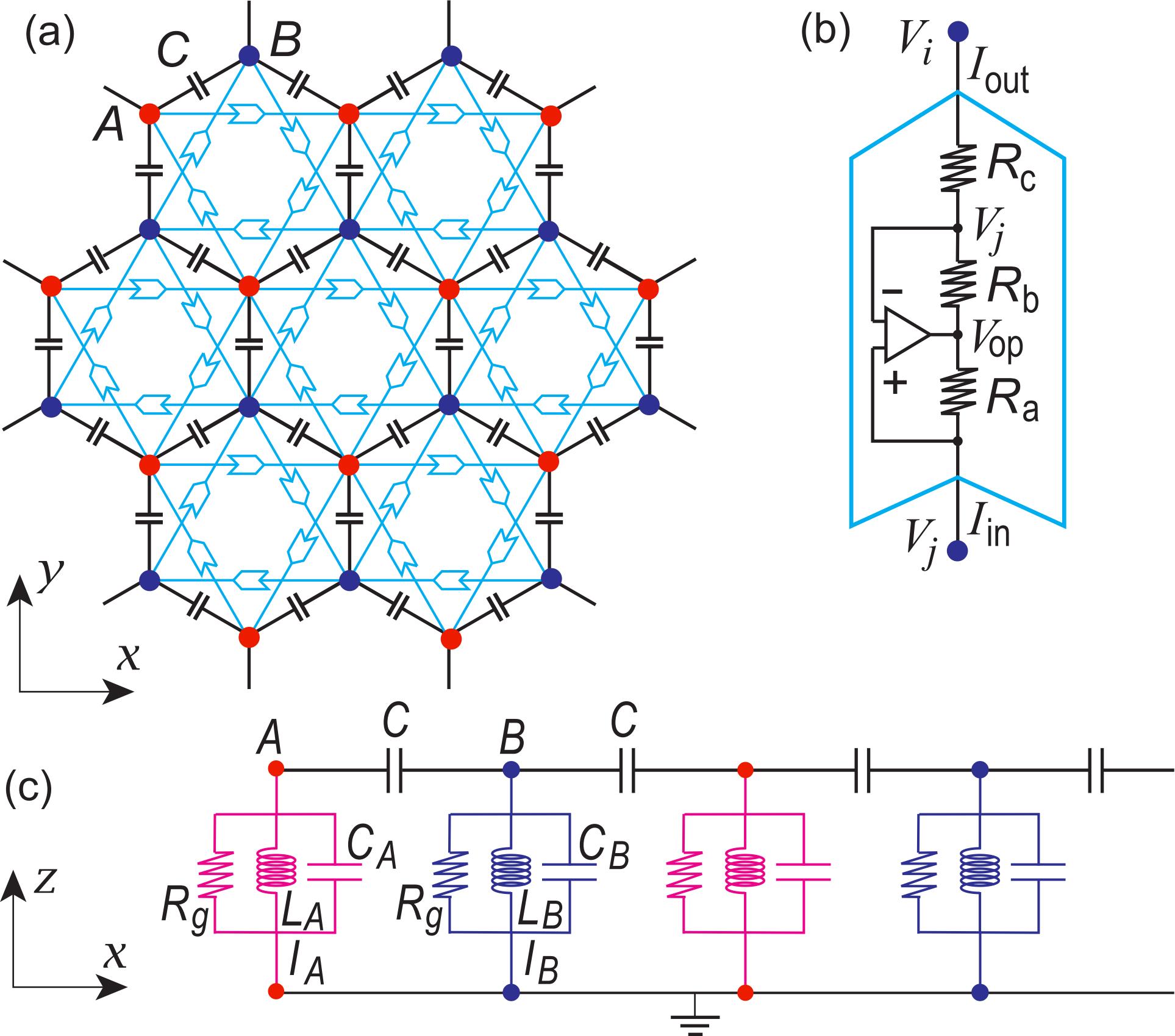}}
\caption{(a) Illustration of an electric circuit realizing the non-Hermitian
Haldane model, which consists of capacitors $C$ and operational amplifiers
acting as the spin-orbit interaction. (b) Structure of an operational
amplifier circuit\protect\cite{Hofmann}. (c) Each node is ground by a set of
capacitor, inductor and resistor ($C_{A},L_{A},R_g$) or ($C_{B},L_{B},R_g$).}
\label{FigCircuit}
\end{figure}

We study a non-Hermitian Haldane model on the honeycomb lattice, where the
spin-orbit interaction is made non-Hermitian. It is constructed by an $LC$
circuit together with operational amplifiers as shown in Fig.\ref{FigCircuit}. 
The spin-orbit term yields complex Dirac masses at the $K$ and $K^{\prime
} $ points. The chiral edge states emerge in the topological phase. They
have imaginary energy. We show a topological phase transition to occur when
the real part of the bulk-energy spectrum is closed. The Chern number is
determined only by the real part of the Dirac mass, while the absolute value
of the bulk energy does not close at the transition point. Non-Hermitian
chiral edges emerge in nanoribbon geometry, which acquire pure imaginary
energy at the high symmetry points. A prominent feature is that the
left-going and right-going chiral edge modes are distinguished by the phase
of the two-point impedance. Furthermore, it is possible to observe only one of
them by a measurement of the one-point impedance.

\textit{Non-Hermitian Haldane model:} We investigate a non-Hermitian Haldane
model on the honeycomb lattice, where the Haldane interaction is
non-Hermitian. As described later, it is realized by an electric circuit of
the configuration illustrated in Fig.\ref{FigCircuit}. The honeycomb lattice
is a bipartite lattice, which consists of the $A$ and $B$ sites. In the
basis of the $A$ and $B$ sites, the Hamiltonian is given by 
\begin{equation}
H=\left( 
\begin{array}{cc}
g\left( \mathbf{k}\right) +U & f\left( \mathbf{k}\right) \\ 
f^{\ast }\left( \mathbf{k}\right) & -g\left( \mathbf{k}\right) -U%
\end{array}%
\right) ,  \label{HamilGraQM}
\end{equation}%
with 
\begin{align}
f\left( \mathbf{k}\right) & =t(e^{-ik_{y}/\sqrt{3}}+2e^{ik_{y}/2\sqrt{3}}\cos \frac{k_{x}}{2}),  \label{FuncF} \\
g\left( \mathbf{k}\right) & =\frac{i}{3\sqrt{3}}[\lambda
^{r}(e^{ik_{x}}+e^{-i\frac{k_{x}+\sqrt{3}k_{y}}{2}}+e^{-i\frac{k_{x}-\sqrt{3}k_{y}}{2}})  \notag \\
& -\lambda ^{l}(e^{-ik_{x}}+e^{i\frac{k_{x}+\sqrt{3}k_{y}}{2}}+e^{i\frac{k_{x}-\sqrt{3}k_{y}}{2}})],
\end{align}%
where $f\left( \mathbf{k}\right) $ describes hopping, and $g\left( \mathbf{k}\right) $ 
describes the non-Hermitian Haldane interaction. It is reduced to
the original Haldane interaction for $\lambda ^{l}=\lambda ^{r}$. We have
added one-site staggered potential $\pm U$, which alternates between $A$ and 
$B$ sites. Let us define $\lambda =\left( \lambda ^{l}+\lambda ^{r}\right)
/2 $ and $\gamma =\left( \lambda ^{l}-\lambda ^{r}\right) /2$, where $\gamma 
$ represents the non-Hermiticity.

\begin{figure}[t]
\centerline{\includegraphics[width=0.48\textwidth]{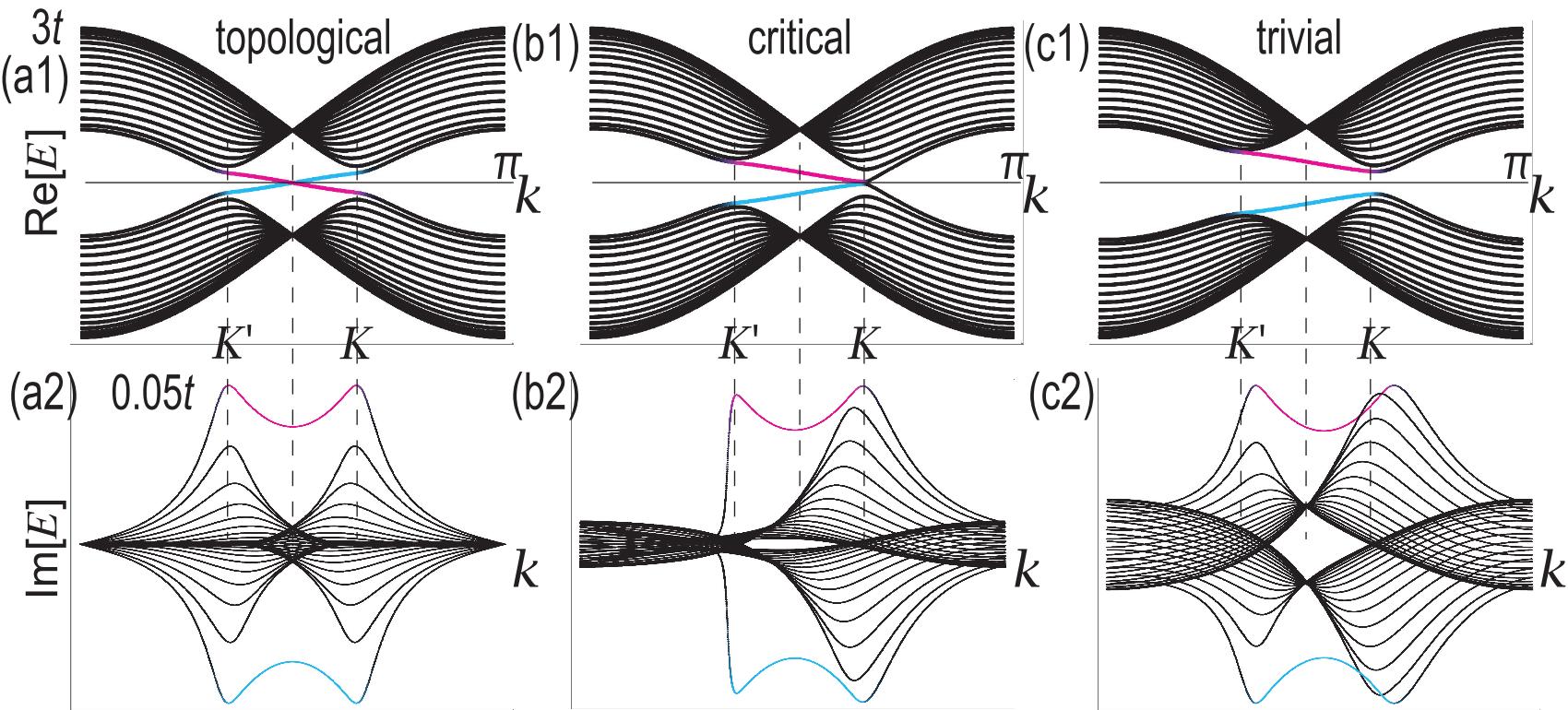}}
\caption{Band structure of the non-Hermitian Chern insulator with zigzag
nanoribbon geometry. (a1)--(c1) real part, and (a2)--(c2) imaginary part of
the band structure. (a) Topological phase ($U=0$), (b) critical point 
($U=\protect\lambda $) and (c) trivial phase ($U=2\protect\lambda $). Color
represents how the wave-function localizes at the right (magenta) or left
(cyan) edges. We have set $\protect\lambda =0.2t$ and $\protect\gamma =0.1t$. }
\label{FigRibbon}
\end{figure}

\begin{figure*}[t]
\centerline{\includegraphics[width=0.92\textwidth]{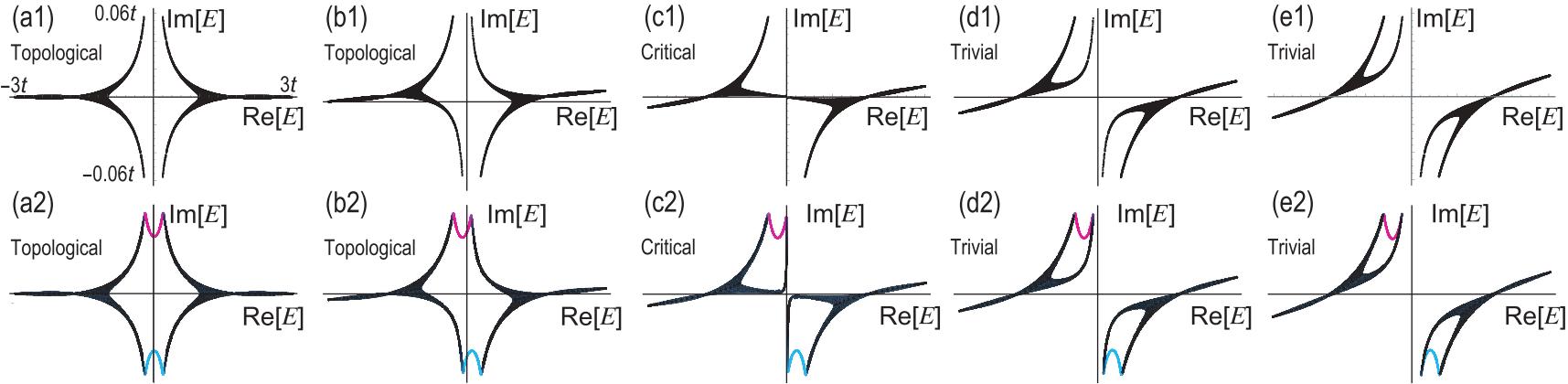}}
\caption{Energy spectrum of the non-Hermitian Chern insulator in Re$E$-Im$E$
plane (a1)--(e1) with bulk and (a2)--(e2) with zigzag nanoribbon geometry.
(a1), (a2) and (b1), (b2) for topological phase ($C=1$) with $U=0$ and $U=0.5\protect\lambda$, respectively; 
(c1), (c2) for critical point with $U=\protect\lambda $; 
(d1), (d2) and (e1), (e2) for trivial phase ($C=0$) with 
$U=1.5\protect\lambda $ and $U=2\protect\lambda $, respectively. Magenta
(cyan) color indicates the edge states localized at the right (left) edge,
while black color indicates the bulk states. The edge states connect the
conduction and valence bands in the topological phase, while they attach
themselves to the same bands in the trivial phase. Parameters 
$\protect\lambda $ and $\protect\gamma $ are the same as in Fig.\protect\ref{FigRibbon}.}
\label{FigChernPlot}
\end{figure*}

The energy is given by $E=\pm \sqrt{\left\vert f\left( \mathbf{k}\right)
\right\vert ^{2}+\left( g\left( \mathbf{k}\right) +U\right) ^{2}}$, which we
show in Fig.\ref{FigRibbon}. Characteristic features read as follows. First
of all, it is complex in general. The real part Re[$E$] looks very similar
to the energy of the original Haldane model with $\gamma =0$, while the
imaginary part Im[$E$] has peaks at the $K$ and $K^{\prime }$ points, i.e.,
at $\mathbf{K}_{\xi }=\left( \xi \frac{4\pi }{3},0\right) $ with $\xi =\pm $. 
The positions of the $K$ and $K^{\prime }$ points do not shift by the
non-Hermitian Haldane interaction. We also show the energy spectrum in the
Re[$E$]-Im[$E$] plane in Fig.\ref{FigChernPlot}(a1)--(e1), where the
conduction and valence bands are separated along the line Re[$E$]=0 except
for $U=\lambda $. It is called a line gap\cite{KawabataST}.

We study physics near the Fermi level. We make the Taylor expansion of (\ref{FuncF}) 
around the $K$ and $K^{\prime }$ points. Hereafter, let us use 
$\mathbf{k}$ as the momentum measured from each of these points. We obtain 
$f\left( \mathbf{k}+\mathbf{K}_{\xi }\right) =\xi k_{x}-ik_{y}$ for $|\mathbf{k|}\ll 1$, 
making the Taylor expansion. Hence, the low-energy physics near
the Fermi level is described by the Dirac theory, 
\begin{equation}
H^{\xi }(\mathbf{k})=\hbar v_{\mathrm{F}}(\xi k_{x}\tau _{x}+k_{y}\tau
_{y})+m_{\xi }\tau _{z},  \label{DiracFree}
\end{equation}%
where $v_{\text{F}}=\frac{\sqrt{3}}{2\hbar }t$ is the Fermi velocity, and 
\begin{equation}
m_{\xi }=U-\xi \lambda -i\gamma /\sqrt{3}  \label{mass}
\end{equation}%
is the Dirac masses at the $K$ and $K^{\prime }$ points.

\textit{Non-Hermitian Chern number:} Since the conduction and valence bands
are separated by the line gap for $U\neq \lambda $, the Chern number is well
defined for $U\neq \lambda $, and characterizes the phase even in the
non-Hermitian theory.

The non-Hermitian Chern number is defined by\cite{Kohmoto,Fu}%
\begin{equation}
C=\frac{1}{2\pi }\int_{\text{BZ}}F\left( \mathbf{k}\right) d^{2}k,
\label{ChernNum}
\end{equation}%
where $F\left( \mathbf{k}\right) $ is the non-Hermitian Berry curvature $%
F\left( \mathbf{k}\right) =\nabla \times A\left( \mathbf{k}\right) $ with
the non-Hermitian Berry connection\cite{Kohmoto,Zhu,Yin,Lieu} $A\left( 
\mathbf{k}\right) =-i\left\langle \psi ^{\text{L}}\right\vert \partial
_{k}\left\vert \psi ^{\text{R}}\right\rangle $. The integration of the Chern
number is taken over the Brillouin zone.

We first summarize general results on the non-Hermitian Chern number in the
two-band systems. We expand the Hamiltonian by the Pauli matrices, 
$H=\sum_{i=x,y,z}h_{i}\tau _{i}$, with $h_{i}$ being complex-valued
functions. Its right and left eigenvalues are given by\cite{Jiang} 
\begin{align}
\left\vert \psi ^{\text{R}}\right\rangle & =[2E\left( E-h_{z}\right)
]^{-1/2}\left( h_{x}-ih_{y},E-h_{z}\right) ^{T},  \notag \\
\left\langle \psi ^{\text{L}}\right\vert & =[2E\left( E-h_{z}\right)
]^{-1/2}\left( h_{x}+ih_{y},E-h_{z}\right) ,
\end{align}%
which satisfy the biorthogonal condition, 
$\left\langle \psi ^{\text{L}}\left\vert \psi ^{\text{R}}\right\rangle \right. =1$. The non-Hermitian
Berry connection is calculated as\cite{Jiang} 
\begin{equation}
A_{\alpha }=(h_{x}\partial k_{k_{\alpha }}h_{y}-h_{y}\partial _{k_{\alpha
}}h_{x})/[E\left( h-E\right) ].
\end{equation}%
The non-Hermitian Berry curvature reads 
\begin{equation}
F\left( \mathbf{k}\right) =2^{-1}E^{-3/2}\varepsilon _{\mu \upsilon \rho
}h_{\mu }\partial _{k_{x}}h_{\nu }\partial _{k_{y}}h_{\rho }.
\end{equation}%
The Chern number (\ref{ChernNum}) with this $F\left( \mathbf{k}\right) $ is
understood as the Pontryagin number or the wrapping number of $h$.

For the $K$ and $K^{\prime }$ points ($\xi =\pm $) we explicitly find 
\begin{equation}
F_{\xi }\left( \mathbf{k}\right) =-\frac{\xi v^{2}}{2}(m_{\xi }+U)[\left(
m_{\xi }+U\right) ^{2}+v^{2}k^{2}]^{-3/2}.
\end{equation}%
As in the case of the Hermitian system\cite{EzawaJPSJReview}, the
integration of $k$ is elongated in (\ref{ChernNum}) from $0$ to $\infty $\
since the Berry curvature has strong peaks at the $K$ and $K^{\prime }$
points. The Chern number is 
\begin{equation}
C_{\xi }=( m_{\xi }+U)/\left( 2\sqrt{\left( m_{\xi }+U\right) ^{2}}\right),
\end{equation}%
which is the same as in the Hermitian system although $m_{\xi }$ now takes a
complex value. We find $C_{\xi }=-\xi /2$ for Re[$m_{\xi }+U$]$>0$ and 
$C_{\xi }=\xi /2$ for Re[$m_{\xi }+U$]$<0$, which are independent of the
non-Hermiticity $\gamma $ since $\gamma $ is the imaginary part of $m_{\xi }$
as in (\ref{mass}). As a result, the topological phase diagram is the same
as in the Hermitian model. The system is topological for $\left\vert \lambda
\right\vert >\left\vert U\right\vert $ and trivial for $\left\vert \lambda
\right\vert <\left\vert U\right\vert $.

\textit{Non-Hermitian chiral edge states:} According to the bulk-edge
correspondence in nanoribbon geometry, the topological phase is
characterized by the emergence of the left-going and right-going chiral edge
states which cross each other at $E=0$. An intuitive reason is that the
topological number cannot change its value without gap closing. We ask how
it is affected by the non-Hermitian Haldane term.

We show the energy spectra in the Re[$E$]-Im[$E$] plane in 
Fig.\ref{FigChernPlot}(a2)--(e2). There are two edge states connecting the valence
and conduction bands in the topological phase. On the other hand, the
edge states attach themselves to the same bands in the trivial phase.

We may understand the structure as follows. In the Haldane model ($\gamma =0$), 
it is well known that the two chiral edges along the two edges cross at 
$E=0$, where $k=\pi $, charactering the topological phase ($m<\lambda $). We
ask how the crossing point moves as the non-Hermiticity $\gamma $ is
introduced. By perturbation theory in $\gamma $, the behavior of $E_{\text{edge}}(k)$ 
at $k=\pi $ is obtained analytically as $E_{\text{edge}}(\pi
)=-2i\lambda \gamma $ for the $A$ sites and $E_{\text{edge}}(\pi )=2i\lambda
\gamma $ for the $B$ sites. Consequently, the crossing points at $k=\pi $
become pure imaginary for $\gamma \neq 0$, as in Fig.\ref{FigChernPlot}(a2).
The bulk-band structure changes suddenly at the critical point ($m=\lambda $), 
resulting in the switching of the edge states between the topological and
trivial phases.

\textit{Electric circuit realization: } The Haldane model is constructed as
in Fig.\ref{FigCircuit}. First, a capacitor $C$ is placed on each
neighboring link of the honeycomb lattice. Second, we bridge a pair of the
same type of sites by an operational amplifier, which acts as a negative
impedance converter with current inversion and act as the spin-orbit
interaction\cite{Hofmann}. Then, $A$ and $B$ sites are grounded 
by a set of capacitor, inductor and resistor ($C_{A},L_{A},R_g$) and ($C_{B},L_{B},R_g$), respectively.

Let us first review\cite{Hofmann} the operational amplifier when it operates
in a negative feedback configuration. The current entering the operational
amplifier from the bottom is given by $I_{\text{in}}=(V_{j}-V_{\text{op}})/R_{a}$, 
while the current leaving from the top is $I_{\text{out}}=(V_{j}-V_{i})/R_{c}$. 
We set an infinite impedance of the operational
amplifier, where any current cannot flow into the operational amplifier.
Then, the output current is given by $I_{\text{out}}=(V_{\text{op}}-V_{j})/R_{b}$. As a result, we find\cite{Hofmann} 
\begin{equation}
\left( 
\begin{array}{c}
I_{\text{in}} \\ 
I_{\text{out}}%
\end{array}%
\right) =\frac{1}{R_{c}}\left( 
\begin{array}{cc}
-\nu & \nu \\ 
-1 & 1%
\end{array}%
\right) \left( 
\begin{array}{c}
V_{j} \\ 
V_{i}%
\end{array}%
\right) ,
\end{equation}%
with $\nu =R_{b}/R_{a}$, where $R_{a}$, $R_{b}$ and $R_{c}$ are the
resistances in the operational amplifier circuit: See Fig.\ref{FigCircuit}(b).

With the AC voltage $V\left( t\right) =V\left( 0\right) e^{i\omega t}$
applied, the Kirchhoff's current law reads\cite{ComPhys,TECNature} 
$I_{a}\left( \omega \right) =\sum_{b}J_{ab}\left( \omega \right) V_{b}\left(
\omega \right) $, where $I_{a}$ is the current between the site $a$ and the
ground, while $V_{a}$ is the voltage at the site $a$. The circuit Laplacian
corresponding to the circuit in Fig.\ref{FigCircuit}(a) is given by
\begin{align}
J\left( \omega \right) & =[3i\omega C+\frac{i\omega C_{A}}{2}+\frac{i\omega
C_{B}}{2}-\frac{1}{2i\omega L_{A}}-\frac{1}{2i\omega L_{B}}+\frac{1}{R_{\text{g}}}]\mathbb{I}  \notag \\
& -\left( 
\begin{array}{cc}
g_{J}\left( \mathbf{k}\right) +U_{J} & f_{J}\left( \mathbf{k}\right) \\ 
f_{J}^{\ast }\left( \mathbf{k}\right) & -g_{J}\left( \mathbf{k}\right) -U_{J}%
\end{array}%
\right) .  \label{Jab}
\end{align}%
with%
\begin{align}
f_{J}\left( \mathbf{k}\right) & =i\omega C\left( e^{-iak_{y}/\sqrt{3}%
}+2e^{iak_{y}/2\sqrt{3}}\cos \frac{ak_{x}}{2}\right) -\frac{1}{i\omega L}, 
\notag \\
g_{J}\left( \mathbf{k}\right) & =\frac{-1}{3\sqrt{3}}[\frac{R_{b}}{R_{a}R_{c}%
}\left( e^{iak_{x}}+e^{-ia\frac{k_{x}+\sqrt{3}k_{y}}{2}}+e^{-ia\frac{k_{x}-%
\sqrt{3}k_{y}}{2}}\right)  \notag \\
& -\frac{1}{R_{c}}\left( e^{-iak_{x}}+e^{ia\frac{k_{x}+\sqrt{3}k_{y}}{2}%
}+e^{ia\frac{k_{x}-\sqrt{3}k_{y}}{2}}\right) ],  \notag \\
U& =\frac{i\omega }{2}(C_{A}-C_{B})-\frac{1}{2i\omega }\left(
L_{A}^{-1}-L_{B}^{-1}\right) .
\end{align}%
Since $J$ is a $2\times 2$ matrix, we may equate it with the Hamiltonian 
(\ref{HamilGraQM}). It follows that $J=i\omega H+1/R_{\text{g}}$, provided
\begin{align}
t& =-C,\quad \lambda ^{r}=R_{b}/(\omega R_{a}R_{c}),\quad \lambda
^{l}=1/\omega R_{c},  \notag \\
U& =-\frac{1}{2}\left( C_{A}-C_{B}\right) +\frac{1}{2\omega ^{2}}\left(
L_{A}^{-1}-L_{B}^{-1}\right) .
\end{align}%
The resistors in the operational amplifier circuit are tuned to be $\nu =1$,
or $R_{a}=R_{b}$, in the literature\cite{Hofmann} so that the system becomes
Hermitian. However, the system is non-Hermitian for a general value of the
resistors, leading to $\nu \neq 1$, which is the main theme of this paper.
It is noted that the topological phase transition is induced solely
controlling the capacitors $C_{A}$, $C_{B}$ or the inductors $L_{A}$, $L_{B}$
connected to the ground. 
We have also added resistors $R_{\text{g}}$ between nodes and the ground, 
whose role is shift the imaginary part of the energy.

\begin{figure}[t]
\centerline{\includegraphics[width=0.48\textwidth]{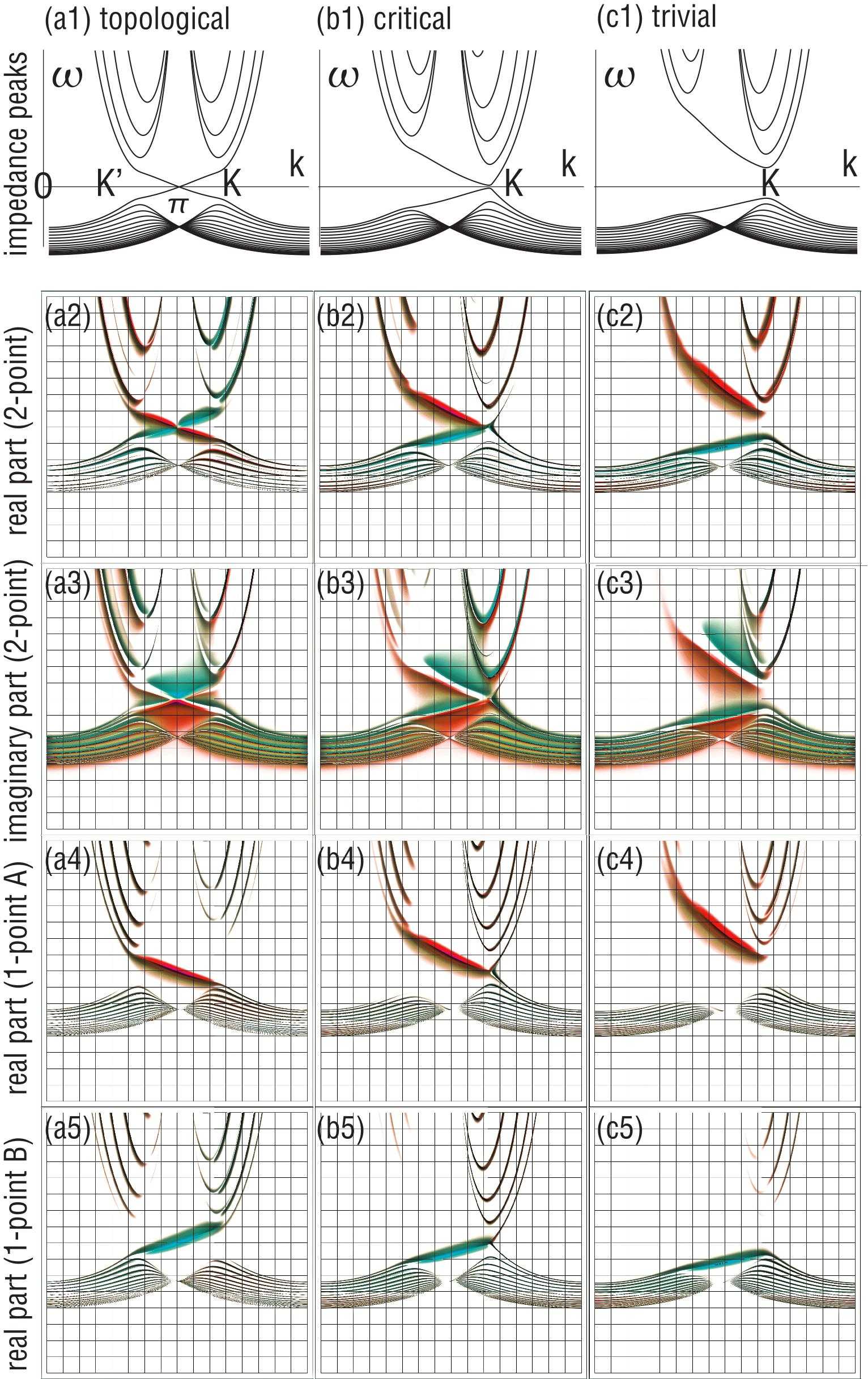}}
\caption{(a1)--(c1) Admittance curves in the $k$-$\protect\omega $ plane
given by (\protect\ref{ResonFrequ}), along which the impedance peaks appear.
(a2)--(c2) Real part of the two-point impedance $Z_{AB}$, which is quite
similar to the impedance peaks in (a1)--(c1). (a3)--(c3) Imaginary part of
the two-point impedance $Z_{AB}$. (a4)--(c4) Real part of the one-point
impedance $G_{A}$ at the edge at the $A$ sites. (a5)--(c5) Real part of the
one-point impedance $G_{B}$ at the edge at the $B$ sites. (a) $U=0$, 
(b) $U=\protect\lambda $ and (c) $U=2\protect\lambda $. The red (green) regions
indicate positive (negative) values of the impedance. 
We have set $\protect\lambda =0.2t$ and $\protect\gamma =0.1t$. }
\label{FigImpe}
\end{figure}

\textit{Admittance spectrum and impedance peaks:} The admittance spectrum
consists of the eigenvalues of the circuit Laplacian\cite%
{ComPhys,TECNature,Garcia,Hel,Lu,EzawaTEC} (\ref{Jab}). 
Indeed, it is possible to observe experimentally the admittance band structure in a direct measurement\cite{Hel}.
It corresponds to the band structure in condensed-matter physics.

Two-point impedance is given by\cite{ComPhys,TECNature} 
$Z_{ab}=\frac{V_{a}-V_{b}}{I_{ab}}=G_{aa}+G_{bb}-G_{ab}-G_{ba}$, where $G$ is the green
function defined by the inverse of the Laplacian $J$, $G\equiv J^{-1}$. In
the case of the Hermitian system, it diverges at the frequency where one of
the eigenvalues of the Laplacian is zero $J_{n}=0$. Namely, the peaks of
impedance are well described by the zeros of the admittance. On the other
hand, in the case of the non-Hermitian system, it takes a maximum value at
the frequency satisfying Re[$J_{n}$]$=0$. After the diagonalization, the
circuit Laplacian yields 
\begin{eqnarray}
J_{n}\left( \omega \right) /i\omega &=&-(2\omega ^{2}L_{A})^{-1}-(2\omega
^{2}L_{B})^{-1}+3C  \notag \\
&&-\varepsilon _{n}\left( \omega \right) +1/i\omega R_{\text{g}},
\end{eqnarray}%
where $\varepsilon _{n}$ is the eigenvalue of the corresponding
Hamiltonian. It is pure imaginary for the Hermitian system but becomes
complex for the non-Hermitian system. Solving Re$J_{n}\left( \omega \right)
=0$, we obtain 
\begin{equation}
\omega _{\text{R}}(\varepsilon _{n})=1/\sqrt{(-\text{Re}\left[ \varepsilon
_{n}\right] +3C)/\left( 1/2L_{A}+1/2L_{B}\right) },  \label{ResonFrequ}
\end{equation}%
at which the impedance has a peak. The eigenvalue $\varepsilon _{n}$ should
be solved numerically. The impedance resonance becomes weaker when the
eigenvalue $\varepsilon _{n}$ has an imaginary part. However, by
tuning the resistors $R_{\text{g}}$, we can decrease the imaginary
part to enhance the impedance peak.

We consider a nanoribbon, where $\omega _{\text{R}}(\varepsilon _{n})$ is a function of the momentum $k_{x}$. 
We show it as a function of $k_{x}$ in Fig.\ref{FigImpe}(a1)--(c1). The impedance has peaks at these
frequencies. The impedance-peak structure in Fig.\ref{FigImpe}(a1)--(c1)
have some feature common to the band structure in Fig.\ref{FigRibbon}(a1)--(c1). 
The chiral edge states cross the $\omega =0$ line at $k=\pi $
when $U=0$ in Fig.\ref{FigImpe}(a1). The crossing point moves towards the $K$
point as $U$ increases and reaches it when $U=\lambda $ as in Fig.\ref{FigImpe}(b1). 
It is the signal of the topological phase. It disappears for $U>0$ as in Fig.\ref{FigImpe}(c1), 
indicating that the system is in the
trivial phase. Thus the topological phase transition induced by the
potential $U$ is clearly observed in the change of the impedance peak.

We calculate the impedance with the use of Green functions as a function of 
$k_{x}$ and $\omega $, and show it in Fig.\ref{FigImpe}(a2)--(c5), where the
impedance strength is represented by darkness. Note that the momentum dependent impedance is an experimentally detectable quantity\cite{Hel,Lee}
by using a Fourier transformation along the nanoribbon direction,%
\begin{equation}
Z_{\alpha \beta }\left( k_{x},y,\omega \right) =\sum_{\rho }Z_{\alpha \beta
}\left( x_{\rho },y_{\rho },\omega \right) \exp \left[ -ix_{\rho }k_{x}%
\right] ,
\end{equation}%
where $\left( x_{\rho },y_{\rho }\right) $ is the Bravais vector.

The impedance is pure imaginary in the Hermitian model. On the other hand,
there emerges the real part of the impedance in the non-Hermitian model,
since the energy becomes complex. It is to be emphasized that the
impedance is a complex object whose real and imaginary parts are separately
observable in electronics by measuring the magnitude and the phase shift of
the impedance. Let us focus on the two-point impedance Re$Z_{AB}$ between
the two outermost edges of a nanoribbon. It is found that the chiral edge
modes become prominent as in Fig.\ref{FigImpe}(a2) and (a3). This is because
the imaginary part of the energy is enhanced at the $K$ and $K^{\prime }$
points. Furthermore, it is a characteristic feature of the non-Hermitian
system that the left-going and right-going chiral edges are differentiated
by the sign of Re$Z_{AB}$ as in Fig.\ref{FigImpe}(a2), where the sign
corresponds to red or green. Equivalently, they are distinguished by the
phase of the impedance. This is because the imaginary part of the energy is
opposite between them.

The one-point impedance is defined by\cite{Hel} $Z_{a}\equiv V_{a}/I_{a}=G_{aa}$. 
It is the inverse of the resistance between the point $a $ and the ground. The one-point impedance $Z_{A}$ at the outermost edge
site $A$ is shown in Fig.\ref{FigImpe}(a4)--(c4), where only the left-going
chiral edge has a strong peak. It is because there is only the left-going
chiral edge at the $A$ edge. On the contrary, there is a strong resonance
only for the right-going chiral edge when we measure $Z_{B}$ as in Fig.\ref{FigImpe}(a5)--(c5). 
This selective detection of the chiral edge is possible
only in electric circuits.

\textit{Discussion:} Topological monolayer systems such as silicene,
germanene and stanene provides us with rich topological phase transitions\cite{EzawaJPSJReview}. 
However, their experimental realizations are yet to
be made. In addition, it is hard to observe the edge states of the two
dimensional topological insulators by ARPES since the intensity is very
weak. On the other hand, almost all of these physics are realizable by
employing electric circuits. In particular, topological phase transitions
are detectable by measuring the change of impedance. Furthermore, it is
possible to observe the left-going and right-going edge states separately by
measuring the one-point impedance.

The author is very grateful to N. Nagaosa for helpful discussions on the
subject. This work is supported by the Grants-in-Aid for Scientific Research
from MEXT KAKENHI (Grants No. JP17K05490, No. JP15H05854 and No.
JP18H03676). This work is also supported by CREST, JST (JPMJCR16F1 and
JPMJCR1874).

\end{document}